\documentstyle[osa,manuscript]{revtex}
\newcommand{\MF}{{\large{\manual META}\-{\manual FONT}}}
\newcommand{\manual}{rm}
\newcommand\bs{\char '134 }
\newcommand{\bsigma}{\mbox{$\boldmath \sigma$}}
\newcommand{\bnabla}{\mbox{$\boldmath \nabla$}}
\newcommand{\bepsilon}{\mbox{$\boldmath \epsilon$}}
\newcommand{\bE}{\mbox{$\boldmath {\cal E}$}}

\begin{document}
\title{The Payne Effect for Particle-Reinforced Elastomers}
\author{Aleksey D. Drozdov$^{1}$ and Al Dorfmann$^{2}$}
\address{$^{1}$ Institute for Industrial Mathematics, 4 Hanachtom Street\\
84311 Beersheba, Israel\\
$^{2}$ Institute of Structural Engineering, 82 Peter Jordan Street\\
1190 Vienna, Austria}
\maketitle

\begin{abstract}
The study deals with the Payne effect (a substantial
decrease in the storage modulus of a particle-reinforced elastomer
with an increase in the amplitude of mechanical oscillations).
The influence of temperature, concentration of filler and
amplitude and frequency of strains is analyzed on the mechanical
response of filled rubbery polymers.
Constitutive equations are derived using the concept
of two interpenetrating networks:
one comprises semiflexible polymeric chains connected
to temporary junctions,
whereas the other is formed by aggregated filler clusters.
Adjustable parameters are found by fitting experimental data
for natural rubber, bromobutyl rubber and styrene--butadiene
rubber reinforced by carbon black and polymeric particles.
The critical concentration of particles is determined that characterizes
transition from an ensemble of disjoint clusters to the network
of filler.
The volume fraction of filler corresponding to this transition is found
to be close to theoretical predictions based on the percolation theory,
as well as to experimental data for isolator--conductor transition.
\end{abstract}

\section{Introduction}

This paper is concerned with the Payne effect
\cite{Pay62,Pay63,Pay65,Pay67}:
a noticeable decrease in the storage modulus of filled rubbery
polymers with an increase in the amplitude of small-strain oscillations
in dynamic mechanical tests.
Changes in dynamic moduli of filled elastomers are
traditionally associated with aggregation of filler particles
into clusters and networks \cite{WRC93,HVH96,HV99,VUH99}.
This phenomenon has attracted substantial attention of experimentalists
in the past decade \cite{VUH99,DKT91,MTD93,LL96,Ulm96,Lio97,Lio98,MDT98,CBY00}
because of its importance for industrial applications (airplane
and automotive tires).

The strain-dependence of dynamic moduli is conventionally described
in terms of the Kraus equations \cite{Kra84}
\begin{equation}
E^{\prime}=E_{0}^{\prime}
+\frac{\Delta E^{\prime}}{1+(\Delta/\Delta_{0})^{2m}},
\qquad
E^{\prime\prime}=E_{0}^{\prime\prime}
+\frac{\Delta E^{\prime\prime}(\Delta/\Delta_{0})^{m}}{1
+(\Delta/\Delta_{0})^{2m}},
\end{equation}
where $E^{\prime}$ and $E^{\prime\prime}$ are storage and loss moduli
in a dynamic test with the strain-amplitude $\Delta$,
$E^{\prime}_{0}$ and $E^{\prime\prime}_{0}$ are dynamic moduli
in a test with an infinitesimally small amplitude of strains,
and $\Delta_{0}$ and $m$ are adjustable parameters.
Equations (1) were developed with reference to the
deagglomeration--reagglomeration mechanism for aggregation
of filler clusters.
The exponent $m$ in Eqs. (1) is fairly well predicted in terms
of the connectivity parameter for fractal aggregates \cite{HVH96}.

Formulas (1) provide an acceptable agreement with experimental data
in dynamic tests with a fixed frequency of oscillations \cite{VUH99,Ulm96}.
However, three shortcomings of the Kraus equations may be mentioned:
\begin{enumerate}
\item
fitting observations for storage and loss moduli separately
results in different values of adjustable parameters \cite{VUH99,Ulm96};

\item
Eqs. (1) do not describe changes in dynamic moduli of
rubbery polymers with temperature, volume fraction of filler and
frequency of oscillations;

\item
the deagglomeration--reagglomeration concept does not distinguish between
the response of an ensemble of disjoint clusters corresponding
to small degrees of loading and that for a filler network
(an agglomerate superstructure \cite{MDT98}) observed in elastomers
with high concentrations of reinforcing particles.
\end{enumerate}
The neglect of the clusters--network transition implies
that parameters of Eq. (1) smoothly change with the filler concentration.
On the other hand, observations in dc conductivity tests
reveal that conductivity of reinforced rubbery polymers
increases by several orders of magnitude when
the volume fraction of filler reaches some percolation threshold
\cite{Car89,Lux93,KS96}.
This contradiction implies the following questions:
(i) whether an analog of the isolator--conductor transition is
observed in dynamic mechanical tests, and (if so), (ii) whether
the threshold concentrations of filler found in mechanical
and dc conductivity experiments are close to one another.

We aim to develop a model which can correctly describe
the effects of amplitude and frequency of oscillations on dynamic moduli.
For this purpose we apply the theory of temporary polymeric networks
\cite{GT46,Yam56,Lod68,TE92} to characterize the time-dependent
behavior of the host material and the concept of bound \cite{HV99,Leb97}
and occluded \cite{WRC93} rubber to predict interactions between
the elastomer and filler particles.

According to conventional theories of transient networks,
a rubbery polymer is thought of as an incompressible network
of long chains connected to temporary junctions.
A chain whose ends are connected to separate junctions
is treated as an active one.
Snapping of an end of a chain from a junction is modeled as
its breakage (transition from its active state to the dangling state).
When a dangling chain captures a junction, a new active chain arises.
Breakage and reformation of active chains (occurring at random times)
are thought of as thermally activated processes.

Unlike previous studies where all chains are treated as composed of an
identical number of strands, we suppose that chains contain various
numbers of strands, $n$, and determine the distribution of chains
by the probability density, $p(n)$, which is assumed to change
with temperature, the volume fraction of filler and the strain intensity.
To reduce the number of adjustable parameters in the governing
equations, we adopt a discrete analog of the Gaussian distribution,
\begin{equation}
p(n)=A \exp \Bigl [-\frac{(n-N)^{2}}{2\Sigma^{2}}\Bigr ]
\qquad
(n=1,\ldots, N_{1}),
\qquad
p(n)=0
\quad
(n=N_{1}+1,\ldots),
\end{equation}
where $N$ is the average number of strands in a chain,
$\Sigma$ is an analog of the standard deviation for the number of strands,
$N_{1}$ is the maximal number of strands in a chain,
and the constant $A$ is determined by the equality
\[ \sum_{n=1}^{N_{1}} p(n)=1. \]
The advantage of Eq. (2) is that given {\it a priori} quantities
$N$ and $N_{1}$, the distribution of chains is determined by the
only parameter $\Sigma$.

Another refinement of the conventional theory of transient networks
is that the rates of breakage and reformation for active chains are assumed
to depend on the number of strands, $n$, which allows relaxation
spectra of filled elastomers to be adequately predicted.

The theories of bound and occluded rubber presume that
macro-strains in the host material are constrained by
surrounding aggregates of filler.
The theory of bound rubber, see Refs. \cite{KS96,Leb97,Med70}
and the references therein, ascribes these constrains to adsorption
of macromolecules on the filler surfaces, which results
in formation of shells of the host material (with the thickness
of about 10 nm \cite{KS96}) around filler particles and their clusters.
Physical properties of the surface layers are assumed to substantially
differ from those for the unfilled elastomer \cite{Wes89}.
In particular, elastic moduli and the resistance to the desorptive
action of a solvent for the shells exceed those for the bulk
polymer by several orders of magnitude \cite{Lux93}.
NRM spectroscopy reveals an internal structure of the shells
which consist of an inner domain of tightly bound macromolecules
and an outer region of loosely bound chains \cite{KS96}.

According to the concept of occluded rubber \cite{WRC93},
aggregation of isolated clusters of particles into a network
results in separation of some meso-regions of the host material
(occluded domains) from the bulk elastomer.
When the filler network is relatively rigid,
this implies shielding of constrained rubbery regions by the network
which does not transmit stresses to the occluded regions.
This results in strong inhomogeneity of macro-deformation.
To simplify the analysis, the host material is assumed to be
split into meso-domains, in a part of which micro-strains
coincide with the macro-strains in the sample, whereas
in the other part micro-strains vanish.

The following scenario is proposed for deformation of
a particle-reinforced elastomer.
The distribution of chains, $p(n)$, in an unfilled rubbery polymer
is characterized by a relatively small standard deviation $\Sigma_{0}$
that determines its relaxation spectrum.
For very small degrees of loading, $\phi$, the parameter $\Sigma$
increases with the growth of the filler concentration
which reflects creation of thin shells containing regions
of loosely bound chains (the number of strands in chains belonging
to these domains is assumed to differ from the ``average'' number for
an unbounded rubber).
For low concentrations of filler, when clusters do not aggregate
into a network [the regime of disjoint clusters (DC)],
changes in the viscoelastic response of a reinforced elastomer
are associated with transformations occurring in surface layers
around the filler particles and their flocks.
We suppose that
\begin{enumerate}
\item
given a temperature, $T$, and an amplitude of oscillations, $\Delta$,
the quantity $\Sigma$ decreases with the degree of loading $\phi$
because of aggregation of isolated particles into clusters,
whose surface area is substantially less than the sum of surface areas of
separated particles;

\item
given a degree of loading, $\phi$, and a strain amplitude, $\Delta$,
the standard deviation of the number of strands, $\Sigma$, increases
with temperature $T$
because of a partial destruction of glassy-like shells of tightly
linked chains around the filler particles and their clusters
and transformation of tightly bound layers into loosely
bound domains;

\item
given a temperature, $T$, and a degree of loading, $\phi$,
the parameter $\Sigma$ weakly decreases with the amplitude
of oscillations $\Delta$
because of mechanically induced breakage of long chains in the regions
of loosely bound rubber and their subsequent amalgamation with the host
elastomer.
\end{enumerate}
At high concentrations of filler, when clusters of reinforcing particles
are aggregated into a network [the regime of network rupture (NR)],
changes in the time-dependent response of a rubbery polymer
are mainly determined by interactions between the host material
and the network.
We suppose that
\begin{enumerate}
\item
given a temperature, $T$, and an amplitude of oscillations, $\Delta$,
the quantity $\Sigma$ strongly decreases with the degree of loading
$\phi$ because of coalescence of flocks into a fractal network,
which causes a noticeable reduction in the lateral surface of
clusters to which regions of loosely bound rubber are adjusted;

\item
given a degree of loading, $\phi$, and a strain amplitude, $\Delta$,
the standard deviation of the number of strands, $\Sigma$, decreases
with temperature $T$
because of the growth of rate of breakage for long chains
in the regions of loosely bound rubber and their subsequent merging
with the bulk polymer;

\item
given a temperature, $T$, and a degree of loading, $\phi$,
the quantity $\Sigma$ dramatically decreases with the amplitude
of oscillations $\Delta$
because of mechanically induced rupture of the filler network
and its crumbling into separated clusters.
\end{enumerate}

The objective of this study is to validate this scenario by comparing
results of numerical simulation with available experimental data.
The exposition is organized as follows.
Section 2 deals with the kinetics of rearrangement for a temporary
network of long chains.
Stress--strain relations for a transient polymeric network are
derived in Section 3.
The constitutive equations are applied to describe uniaxial extension
of a specimen in Section 4.
Section 5 is concerned with comparison of experimental data for
filled rubbery polymers with results of numerical simulation.
Some concluding remarks are formulated in Section 6.

\section{Rearrangement of a temporary network}

The distribution of chains in a polymeric network is determined by the total
number of active chains per unit mass, $\Xi$,
and the probability density of long chains composed of $n$ strands,
$p(n)$.
We focus on the stationary response of filled elastomers
in dynamic mechanical tests with small strains and assume
that the quantities $\Xi$ and $p(n)$ are independent of time
(however, these parameters may, in general, depend on the amplitude
of oscillations).

Denote by $X(t,\tau,n)$ the number of active chains with $n$ strands
(per unit mass) at time $t$ that has last been linked to the network at
instant $\tau\in [0,t]$.
The function $X(t,\tau,n)$ entirely determines the current state
of a polymeric network:
\begin{equation}
X(t,t,n)=\Xi p(n)
\end{equation}
equals the number of active chains (per unit mass) with $n$ strands
at time $t$;
\[ \frac{\partial X}{\partial \tau} (t,\tau,n)\biggl |_{t=\tau} \; d\tau \]
is the number of active chains with $n$ strands (per unit mass)
that have merged with the network within the interval
$[\tau,\tau+d\tau ]$;
\[ \frac{\partial X}{\partial \tau} (t,\tau,n)\;d\tau \]
is the number of these chains that did not break during the
interval $[\tau, t]$;
\[ -\frac{\partial X}{\partial t} (t,0,n)\;dt \]
is the number of active chains (per unit mass) that detach from the
network (for the first time) within the interval $[t,t+dt ]$,
and
\[ -\frac{\partial^{2} X}{\partial t\partial \tau} (t,\tau,n)\;dtd\tau \]
is the number of chains (per unit mass) that has last been linked
to the network within the interval $[\tau,\tau+d\tau ]$
and detach from the network (for the first time after merging)
during the interval $[t,t+dt ]$.

The kinetics of reformation is determined by the relative rate
of breakage for active chains, $\Gamma(n)$,
and by the rate of merging of dangling links with the
network, $\gamma(n)$.
The relative rate of breakage is defined as the ratio of the number
of active chains broken per unit time to the total number of active chains,
\begin{equation}
\Gamma(n) = -\frac{1}{X(t,0,n)}
\frac{\partial X}{\partial t}(t,0,n)=
-\Bigl [ \frac{\partial X}{\partial \tau}
(t,\tau,n) \Bigr ]^{-1}
\frac{\partial^{2} X}{\partial t\partial \tau}(t,\tau,n).
\end{equation}
The rate of breakage $\Gamma$ may, in general, depend on the current
time $t$, on the instant $\tau$ when a chain has been last bridged to
the network and on the guiding vector ${\bf l}$
(the unit vector whose direction coincides with the end-to-end
vector ${\bf R}$ of the chain at time $\tau$).
The neglect of these dependences means that the network is homogeneous
(the rate of breakage for a chain is independent of the instant
of its connection to the network, $\tau$, and on the current time, $t$)
and isotropic (the rate of breakage is independent of the chain direction
at the instant of merging).
For a homogeneous and isotropic network, the rate of merging, $\gamma$,
is independent of the instants $t$ and $\tau$, as well as of
the vector ${\bf l}$.
It is defined as the number of dangling chains (per unit mass)
bridged to the network per unit time,
\begin{equation}
\gamma(n)=\frac{\partial X}{\partial \tau} (t,\tau,n)\biggl |_{\tau=t}.
\end{equation}
Integration of differential equations (4) with the initial conditions (3)
and (5) implies that
\begin{equation}
X(t,0,n) = \Xi p(n) \exp \Bigl [ -\Gamma(n)t \Bigr ],
\qquad
\frac{\partial X}{\partial \tau}(t,\tau,n) =
\gamma(n)\exp \Bigl [-\Gamma(n)(t-\tau) \Bigr ].
\end{equation}
Because the total number of active chains remains time-independent,
the number of broken chains coincides with the number of reformed chains
\[ \gamma(n)=-\frac{\partial X}{\partial t}(t,0,n)
-\int_{0}^{t} \frac{\partial^{2}X}{\partial t\partial \tau}(t,\tau,n)
d\tau. \]
Substitution of Eq. (6) into this equality
implies the balance law
\[ \gamma(n)=\Gamma(n)\biggl \{ \Xi p(n)\exp \Bigl [-\Gamma(n)t\Bigr ]
+\gamma(n)\int_{0}^{t} \exp \Bigl [-\Gamma(n)(t-\tau)\Bigr ]d\tau \biggr \}.
\]
The solution of this equation reads
\begin{equation}
\gamma(n)=\Xi \Gamma(n) p(n).
\end{equation}
Equations (6) and (7) entirely determine the current state for a
temporary network of long chains.

\section{Constitutive relations for a transient network}

The stress in a dangling chain is assumed to totally relax
before this chain captures a new junction, which implies that
the natural (stress-free) state of an active chain coincides with
the deformed state of the network at the instant of their merging.
The extension ratio for a chain bridged to the network at time $\tau$
and not broken within the interval $[\tau, t]$ reads \cite{Dro98}
\[ \lambda(t,\tau,{\bf l})
=[ {\bf l}\cdot{\bf C}(t,\tau)\cdot{\bf l} ]^{\frac{1}{2}}, \]
where ${\bf C}(t,\tau)$ is the relative Cauchy deformation tensor
for transition from the deformed state of the network
at time $\tau$ to the deformed state at time $t$
and the dot stands for inner product.
At small strains, this equality implies that
\begin{equation}
\varepsilon(t,\tau,{\bf l})={\bf l}\cdot\bE (t,\tau)\cdot{\bf l},
\qquad
\bE (t,\tau)=\bepsilon (t)-\bepsilon (\tau),
\end{equation}
where $\varepsilon=\lambda-1$ is the longitudinal strain for a chain,
$\bE =\frac{1}{2}({\bf C}-{\bf I})$ is the relative
Cauchy strain tensor for the network, ${\bf I}$ is the unit tensor
and $\bepsilon (t)$ is the strain tensor for transition from
the stress-free state of the network to its deformed state at time $t$.
By analogy with the Kratky--Porod model, we suppose that strains
in all strands coincide, which implies that the strain
per strand equals $\varepsilon/n$ for a chain comprising $n$ strands.
Modeling strands as linear elastic solids with a constant rigidity $\mu$,
we find the mechanical energy of a chain,
\[ \frac{1}{2} \sum_{i=1}^{n} \mu \Bigl (\frac{\varepsilon}{n}\Bigr )^{2}
=\mu\frac{\varepsilon^{2}}{2n}. \]
With reference to the conventional assumption that
the excluded-volume effect
and other multi-chain effects are screened for an individual chain
by surrounding macromolecules,
we neglect the energy of interaction between chains
and calculate the mechanical energy of a network as the sum of
the mechanical energies for individual chains.
The mechanical energy (per unit mass) of initial chains
which have not been broken within the interval $[0,t]$ is given by
\begin{equation}
\frac{1}{4\pi} \sum_{n=1}^{\infty} \frac{\mu}{2n} X(t,0,n)
\int_{\cal S} \varepsilon^{2}(t,0,{\bf l}) dA({\bf l}),
\end{equation}
where ${\cal S}$ is the unit sphere in the space of vectors ${\bf l}$
and $dA({\bf l})$ is the surface element on ${\cal S}$.
The mechanical energy (per unit mass) of chains
that merged with the network within the interval
$[\tau,\tau+d\tau ]$ and are connected with the network until
the current time $t$ reads
\begin{equation}
\frac{1}{4\pi} \sum_{n=1}^{\infty} \frac{\mu}{2n}
\frac{\partial X}{\partial \tau}(t,\tau,n) d\tau
\int_{\cal S} \varepsilon^{2}(t,\tau,{\bf l}) dA({\bf l}).
\end{equation}
To calculate the integrals over ${\cal S}$, we introduce a Cartesian
coordinate frame $\{ x_{i} \}$ whose base vectors ${\bf k}_{i}$ are
directed along the eigenvectors of the symmetrical tensor $\bE$:
\[ \bE={\cal E}_{1}{\bf k}_{1} +{\cal E}_{2}{\bf k}_{2}
+{\cal E}_{3}{\bf k}_{3}, \]
where ${\cal E}_{i}$ is the $i$th eigenvalue of the tensor $\bE$.
The position of the unit vector ${\bf l}$ with respect to the coordinate
frame is determined by the Euler angles $\varphi$ and $\vartheta$,
\[ {\bf l}=\sin\vartheta (\cos\varphi\;{\bf k}_{1}
+\sin\varphi\;{\bf k}_{2})+\cos\vartheta\;{\bf k}_{3}. \]
Substitution of these expressions into Eq. (8) results in
\[ \varepsilon=\sin^{2}\vartheta ({\cal E}_{1}\cos^{2}\varphi+
{\cal E}_{2}\sin^{2}\varphi)+{\cal E}_{3}\cos^{2}\vartheta. \]
Bearing in mind that $dA=\sin\vartheta d\vartheta d\varphi$,
we obtain
\begin{eqnarray*}
\int_{\cal S} \varepsilon^{2} dA &=& \int_{0}^{\pi}\sin\vartheta d\vartheta
\int_{0}^{2\pi} \Bigl [ \sin^{2}\vartheta ({\cal E}_{1}\cos^{2}\varphi
+{\cal E}_{2}\sin^{2}\varphi)+{\cal E}_{3}\cos^{2}\vartheta \Bigr ]^{2}
d\varphi
\nonumber\\
&=& \frac{4\pi}{15} \Bigl [ 3({\cal E}_{1}^{2}+{\cal E}_{2}^{2}
+{\cal E}_{3}^{2})+2({\cal E}_{1}{\cal E}_{2}
+{\cal E}_{1}{\cal E}_{3}+{\cal E}_{2}{\cal E}_{3})\Bigr ].
\end{eqnarray*}
Summing expressions (9) and (10) and using this equality, we arrive
at the formula for the strain energy density of the network (per unit mass)
\begin{eqnarray*}
U(t) &=& \frac{\mu}{30} \sum_{n=1}^{\infty}  \biggl \{ X(t,0,n)
\Bigl [ 3I_{1}^{2}(\bE (t,0)) -4I_{2}(\bE (t,0))\Bigr ]
\nonumber\\
&&+ \int_{0}^{t} \frac{\partial X}{\partial \tau}(t,\tau,n)
\Bigl [ 3I_{1}^{2}(\bE (t,\tau))
-4I_{2}(\bE (t,\tau))\Bigr ] d\tau \biggr \},
\end{eqnarray*}
where $I_{i}$ stands for the $i$th principal invariant of a tensor.
It follows from this equality, the incompressibility condition,
$I_{1}(\bE )=0$, and Eq. (8) that
\begin{equation}
U(t) = -\frac{2\mu}{15} \sum_{n=1}^{\infty}  \frac{1}{n} \biggl [ X(t,0,n)
I_{2}\Bigl (\bepsilon _{\rm d}(t)\Bigr )
+\int_{0}^{t} \frac{\partial X}{\partial \tau}(t,\tau,n)
I_{2}\Bigl (\bepsilon _{\rm d}(t)-\bepsilon _{\rm d}(\tau)\Bigr )
d\tau \biggr ],
\end{equation}
where $\bepsilon _{\rm d}$ is the deviatoric part of the strain
tensor $\bepsilon $.
The deviatoric component $\bsigma _{\rm d}$ of the stress tensor
$\bsigma $ is determined by the conventional relationship
\[ \bsigma _{\rm d}(t)
=\rho \frac{\partial U(t)}{\partial \bepsilon _{\rm d}(t)}, \]
where $\rho$ is mass density.
Substitution of Eqs. (6), (7) and (11) into this equality results in
the stress--strain relation
\begin{eqnarray}
\bsigma (t) &=& -P(t){\bf I}
+\frac{2}{15}\mu \rho \Xi \sum_{n=1}^{\infty}  \frac{p(n)}{n}
\biggl [ \exp \Bigl ( -\Gamma(n)t\Bigr )\bepsilon _{\rm d}(t)
\nonumber\\
&& +\Gamma(n) \int_{0}^{t} \exp \Bigl (-\Gamma(n) (t-\tau)\Bigr )
\Bigl (\bepsilon _{\rm d}(t)-\bepsilon _{\rm d}(\tau)\Bigr )
d\tau \biggr ],
\end{eqnarray}
where $P$ is pressure.

\section{Uniaxial extension of a network}

For uniaxial extension of an incompressible network
along the axis $x_{1}$ of the Cartesian coordinate frame $\{ x_{i} \}$,
the strain tensor reads
\[ \bepsilon (t)=\epsilon(t) [ {\bf k}_{1}{\bf k}_{1}
-\frac{1}{2} ({\bf k}_{2}{\bf k}_{2}+{\bf k}_{3}{\bf k}_{3})], \]
where $\epsilon(t)$ is the longitudinal strain.
Substitution of this equality into Eq. (12) results in the formula
for the longitudinal stress $\sigma(t)$,
\begin{equation}
\sigma(t) = \frac{1}{5}\mu \rho \Xi \sum_{n=1}^{\infty}  \frac{p(n)}{n}
\biggl [ \epsilon(t)
-\Gamma(n) \int_{0}^{t} \exp \Bigl (-\Gamma(n) (t-\tau)\Bigr )
\epsilon(\tau) d\tau \biggr ].
\end{equation}
We now aim to established a correspondence between the longitudinal strain
for the network, $\epsilon(t)$, and the macro-strain for a specimen,
\begin{equation}
e(t)=\frac{L(t)-L(0)}{L(0)},
\end{equation}
where $L(t)$ is the current length of a specimen.
At low concentrations of filler (the DC regime),
we assume that the host material is homogeneous
and postulate that the micro-strains in active chains coincide
with the macro-strain in the specimen,
\begin{equation}
\epsilon(t)=e(t).
\end{equation}
At high degrees of loading (the NR regime), deformation of a part
of the rubbery polymer is assumed to be screened by the network.
To describe restrictions imposed by the filler network
on deformation of a reinforced elastomer, we (i) postulate that
the network is rigid and (ii) characterize the level of constrains
by the volume fraction of the undeformed host material.
For uniaxial loading, this fraction is uniquely determined
by the length, $L_{0}(t)$, of a part of the sample occupied
by occluded rubber.
Under the hypothesis that deformation in the rest of the specimen
is homogeneous, we find the longitudinal elongation for the
unconstrained part of the bulk material as the ratio of the current
length of this part to its initial length,
\[ \lambda_{\ast}(t)=\frac{L(t)-L_{0}(t)}{L(0)-L_{0}(t)}. \]
We substitute Eq. (14) into this equality,
introduce the longitudinal strain by the conventional formula,
$e_{\ast}=\lambda_{\ast}-1$,
neglect terms of the second order of smallness compared with $e(t)$,
and postulate that the micro-strains in long chains coincide with
the macro-strain in the unrestricted part of the rubbery polymer.
This implies that
\begin{equation}
\epsilon(t) =\frac{L(0)}{L(0)-L_{0}(t)}e(t).
\end{equation}
When the filler network is absent, $L_{0}(t)=0$, and Eq. (16) turns
into Eq. (15).
A similar equality was proposed in Ref. \cite{MT65}
based on the concept of the strain-amplification factor.
Unlike previous studies, where the coefficient on the right-hand side
of Eq. (16) was expressed in terms of the filler concentration
and the ratio of the average length to the average width of chain-like
filler clusters by using a phenomenological equation,
we do not concretize the shape of the function $L_{0}(t)$
and determine it by fitting observations.
As will be demonstrated in Section 5, this function strongly
depends on the strain intensity, which implies that treatment
of $L_{0}$ as a function of the volume fraction of filler
is an oversimplification.

For a dynamic test with an amplitude $\Delta$ and a frequency $\omega$,
we set
\begin{equation}
e(t)=\Delta \exp ({\rm i}\omega t),
\end{equation}
where ${\rm i}=\sqrt{-1}$.
Combining Eqs. (13), (16) and (17) and introducing
the new variable $s=t-\tau$, we find that
\[ E^{\ast}(t,\omega)=K_{0}(t) \sum_{n=1}^{\infty}
\frac{p(n)}{n}\biggl [ 1-\Gamma(n)\int_{0}^{t}
\exp \Bigl (-\Bigl (\Gamma(n)+{\rm i}\omega\Bigr )s\Bigr )
\frac{L(0)-L_{0}(t)}{L(0)-L_{0}(t-s)} ds\biggr ], \]
where
\[ E^{\ast}(t,\omega)=\frac{\sigma(t)}{e(t)} \]
is the complex modulus and
$K_{0}(t)=\frac{1}{5}\mu \rho \Xi L(0)[ L(0)-L_{0}(t) ]^{-1}$.

It is natural to assume that under periodic loading, the function
$L_{0}(t)$ rapidly approaches its limiting value $L_{\infty}(\Delta)$
corresponding to the steady-state regime of oscillations.
Replacing the function $L_{0}(t)$ by the constant $L_{\infty}(\Delta)$,
putting $t=\infty$ as the upper limit of integration,
and splitting $E^{\ast}$ into the sum
\[ E^{\ast}=E^{\prime}+{\rm i}E^{\prime\prime}, \]
where $E^{\prime}(\omega)$ is the storage modulus and
$E^{\prime\prime}(\omega)$ is the loss modulus,
we arrive at the formulas
\begin{equation}
E^{\prime}(\omega) = K\omega^{2}\sum_{n=1}^{\infty}
\frac{p(n)}{n[\Gamma^{2}(n)+\omega^{2}]},
\quad
E^{\prime\prime}(\omega) = K\omega\sum_{n=1}^{\infty}
\frac{p(n)\Gamma(n) }{n[\Gamma^{2}(n)+\omega^{2}]}
\end{equation}
with
\begin{equation}
K=\frac{1}{5}\mu \rho \Xi \frac{L(0)}{L(0)-L_{\infty}}.
\end{equation}

Given quantities $N$ and $N_{1}$, Eqs. (18) are determined by two
adjustable parameters, $K$ and $\Sigma$, and
the rate of slippage from temporary junctions, $\Gamma(n)$.
To reduce the number of constants to be found, we set
\begin{equation}
\Gamma(n)=\Gamma_{0} n^{\kappa},
\end{equation}
where $\Gamma_{0}$ and $\kappa$ are positive parameters.
To explain the increase in the rate of breakage for active
chains with the number of strands, we assume that chains
have finite bending stiffness.
A semiflexible chain is treated as a curvilinear rod whose micro-motion
is confined to some tube formed by surrounding macromolecules \cite{DE86}.
When the rod has a negligible bending rigidity, lateral oscillations
driven by thermal fluctuations have a local character,
which implies that slippage of a chain end from a junction
is induced by random excitations occurring (rather sparsely)
in the close vicinity of this end.
For a rod with a finite bending rigidity,
lateral oscillations induce transverse waves,
whose amplitudes increase because of their interaction with end-points.
The amplification of random transverse oscillations of an elastic rod
in the vicinity of its end-points results in the growth of
the rate of slippage of chain ends from temporary junction.
The account for finite bending stiffness of semiflexible chains
allows the monotonic growth of the function $\Gamma(n)$ to
be explained.
It does not, however, result in the power-law dependence (20),
which is treated as purely phenomenological.

\section{Comparison with experimental data}

We begin with fitting observations in dynamic tensile tests for
a carbon black (CB) filled natural rubber vulcanizate
with the degree of loading $\phi=50$ phr (parts per hundred parts of rubber)
at various temperatures $T$.
For a detailed description of specimens and the experimental procedure,
see Ref. \cite{Lio97}.
To approximate experimental data, we set $N=40$, which is in qualitative
agreement with the average number of strands used in other studies
(for example, $N=12$ was assumed for glassy styrene-acrylonitrile
polymer \cite{Ges97} and $N=20$ was employed in molecular dynamics
simulations of interactions between a polymeric melt and
filler particles \cite{SSG00}), and $N_{1}=500$.
Numerical analysis demonstrates that changes in the values
of $N$ and $N_{1}$ (in the range from $N=30$ to $N=50$
and from $N_{1}=100$ to $N_{1}=1000$) weakly affect the quality
of fitting.

At any temperature $T$, we, first, fit the curves $E^{\prime}(\omega)$
measured at the minimum amplitude of oscillations,
$\Delta_{\min}=0.006$, and determine parameters $\Gamma_{0}$,
$\kappa$ and $\Sigma$ (by the steepest-descent procedure)
which ensure the best quality of matching.
Afterwards, we fix the values of these quantities and approximate
observations by using only one adjustable parameter, $K$,
which is determined by the least-squares algorithm.
Figures~1 demonstrate fair agreement between experimental data and
results of numerical simulation.

The independence of the standard deviation of the number
of strands, $\Sigma$, from the amplitude of oscillations, $\Delta$,
implies that the viscoelastic response of the filled rubber
corresponds to the DC regime (low degrees of loading)
at all temperatures $T$.

The elastic modulus $K$ is plotted versus the amplitude
of oscillations $\Delta$ in Figure~2A which demonstrates
that the modulus slowly decreases with the strain amplitude.
Experimental data are fairly well approximated by the function
\begin{equation}
\log K=a_{0}-a_{1}\Delta,
\end{equation}
where adjustable parameters $a_{i}$ are found by the least-squares
technique.
According to Eq. (19), below the cluster--network transition point,
the modulus $K$ is proportional to the number of active chains
per unit mass, $\Xi$.
A decrease in $K$ with $\Delta$ means that the growth of strains
results in a reduction of the number of active chains because of
the mechanically induced agitation of the breakage process.
This conclusion is in agreement with observations for unfilled rubbery
polymers \cite{Dro98}.

The quantities $\Sigma$ and $a_{0}=\log K(0)$ (the logarithm of
the elastic modulus $K$ for the infinitesimally small amplitude
of oscillations) are plotted in Figure~2B versus the degree
of undercooling, $\Delta T=T-T_{\rm g}$, where $T_{\rm g}$ is
the glass transition temperature.
Experimental data are approximated by the functions
\begin{equation}
\log \Sigma=b_{0}-\frac{b_{1}}{\Delta T},
\qquad
\log K(0)=c_{0}+\frac{c_{1}}{\Delta T},
\end{equation}
where the parameters $b_{i}$ and $c_{i}$ are found by the least-squares
algorithm.
Equations (22) provide good fitting of observations.
An increase in the standard deviation of the number of strands $\Sigma$
with temperature is in agreement with the proposed mechanism for
transformation of the tightly bound domains into the loosely bound
regions.
A decrease in the elastic modulus with temperature confirms the
hypothesis about the elastic nature of the response of long chains
(according to the conventional theory of entropic elasticity,
the modulus linearly increases with temperature because of the
growth in the number of available configurations
for long chains \cite{Tre75}).
Equations (19) and (22) imply that the rigidity of strands, $\mu$,
decreases with temperature.

The rate of breakage for active chains $\Gamma_{0}$ is plotted versus
the degree of undercooling in Figure~2C, which demonstrates
that $\Gamma_{0}$ monotonically decreases with temperature
and reaches its limiting value, $\Gamma_{\infty}$, at high temperatures.
Experimental data are approximated by the phenomenological relation
\begin{equation}
\log \Gamma_{0}=d_{0}-d_{1} \Delta T
\end{equation}
with adjustable parameters $d_{i}$.

At first glance, the decrease in the rate of breakage with temperature
is rather surprising, because the growth of temperature should result
in an increase in the rate of thermally activated processes.
Observations depicted in Figure~2C may be adequately explained,
however, in the framework of the model of transient networks
of semiflexible chains.
The growth of temperature implies, on the one hand, an increase
in the rate of thermal fluctuations which intensifies the breakage
process.
On the other hand, it induces a decrease in the bending rigidity of
elastic chains, which means that transverse oscillations
driven by thermal fluctuations at intermediate points of a chain
are not transmitted to its ends, and, as a consequence,
their amplitudes do not grow because of interactions of bending waves
with the end-points.

The parameter $\kappa$ that describes the influence of the length
of an active chain on its rate of rearrangement is depicted in
Figure~2D as a function of temperature $T$.
The quantity $\kappa$ increases with $T$ at relatively low
temperatures and becomes practically constant at elevated temperatures.

We proceed with fitting observations in dynamic tensile tests
at room temperature for CB filled bromobutyl rubber.
A detailed description of specimens and the experimental
procedure is given in Ref. \cite{DKT91}.
To reduce the number of adjustable parameters,
we fix the average number of strands in a chain
(the same as for the CB filled natural rubber), $N=40$,
and set $N_{1}=80$ (which implies the symmetry of the probability density
$p(n)$ with respect to the point $N$).
Matching observations in a tensile test with the minimum amplitude
of oscillations $\Delta_{\min}=0.0003$ results in
$\Gamma_{0}=0.0088$ s$^{-1}$.
We fix this value and approximate experimental data in other tests
with the help of the adjustable parameters $\kappa$, $\Sigma$ and $K$.
For any degree of loading, $\phi$, we determine the parameter $\kappa$
by fitting the curve $E^{\prime}(\omega)$ measured at the
minimum amplitude of strains, $\Delta_{\min}$.
Afterwards, we fix the value of $\kappa$ and match observations in
dynamic tests with other amplitudes of oscillations by using only
two adjustable parameters, $\Sigma$ and $K$.
The quantity $\Sigma$ is determined using the steepest-descent
procedure and the constant $K$ is found by the least-squares
algorithm.
Figure~3 demonstrates fair agreement between experimental data and
results of numerical simulation.

The parameters $\Sigma$ and $K$ are plotted versus the
amplitude of oscillations $\Delta$ in Figure~4.
Experimental data are fairly well approximated by the functions
\begin{equation}
\log \Sigma=A_{0}-A_{1}\log \Delta,
\qquad
\log K=B_{0}+B_{1}\log \Delta,
\end{equation}
where the parameters $A_{i}$ and $B_{i}$ are determined using
the least-squares algorithm.

Two different regimes in the material behavior are revealed.
At small amplitudes, $\Delta<\Delta_{\rm cr}$,
$\Sigma$ and $K$ weakly change with $\Delta$ (curves~1).
The standard deviation of the number of strands decreases at
the minimum degree of loading, $\phi=30$ phr,
and remains practically constant for $\phi\geq 50$ phr.
The elastic modulus $K$ increases with $\Delta$
at $\phi\leq 50$ phr and decreases at $\phi > 50$ phr.
These changes may be associated with transition of bound rubber
from the tight state to the loose state.
At large amplitudes, $\Delta>\Delta_{\rm cr}$,
$\Sigma$ monotonically decreases and $K$ strongly increases
with $\Delta$ (curves~2), which may be explained by
disintegration of the filler network and release of occluded rubber.

The parameters $A_{0}$ and $A_{1}$ corresponding to the post-fracture
regime for the filler network are plotted versus the degree of loading
$\phi$ in Figure~5A.
Analogous dependences for the sub-fracture regime are not presented
because of the insufficient number of experimental data for the
parameter $\Sigma$ (the best fit of observations depicted in Figures~4C
and 4D is reached for the homogeneous distribution of chains with
various numbers of strands).
Experimental data in Figure~5A are fairly well approximated by the
functions
\begin{equation}
A_{0}=A_{00}+A_{01}\phi,
\qquad
A_{1}=A_{10}-A_{11}\phi
\end{equation}
where adjustable parameters $A_{ij}$ are found by the least-squares
method.
It follows from Eqs. (24) and (25) that in the interval of strains
under consideration, the standard deviation of the number
of strands decreases with the filler concentration,
in agreement with the proposed scenario (at high volume fractions
of filler, aggregation of clusters into a network results in a decrease
in the number of isolated particles, and, as a consequence,
in a decrease in the number of chains belonging to the surface layers).

The parameter $B_{0}$ and $B_{1}$ for the sub-fracture regime for
the filler network are depicted in Figure~5B versus the degree of
loading $\phi$.
Experimental data are correctly predicted by the functions
\begin{equation}
B_{0}=B_{00}-B_{01}\phi,
\qquad
B_{1}=B_{10}-B_{11}\phi,
\end{equation}
where adjustable parameters $B_{ij}$ are determined by the least-squares
algorithm.
Because the quantities $B_{0}$ and $B_{1}$ remain practically constant
in the post-fracture regime, appropriate figures for these parameters
are omitted.

Results presented in Figures~5A and 5B may be treated as a confirmation
of the model because they reveal a plausible (from the physical
standpoint) behavior of adjustable parameters in the constitutive
equations with changes in the filler concentration.

The critical amplitude of oscillations $\Delta_{\rm cr}$
corresponding to fracture of the filler network is plotted
in Figure~5C versus the degree of loading $\phi$.
Observations are fairly well approximated by the function
\begin{equation}
\Delta_{\rm cr}=C(\phi-\phi_{\rm cr})^{-\beta},
\end{equation}
where the parameters $C$, $\beta$ and $\phi_{\rm cr}$ are found by
the steepest-descent procedure.
Following Ref. \cite{VUH99}, we find that the critical
loading $\phi_{\rm cr}=41$ prh corresponds to the critical volume
fraction $\psi_{\rm cr}=0.154$, which is rather close to the
theoretical percolation threshold $\psi_{\ast}=0.18$ for a spatial
network of hard spheres \cite{PS74}, as well as to the critical
volume fraction $\psi_{\star}=0.14$ for the isolator--conductor transition
for CB natural rubber determined in dc conductivity tests \cite{Car89}.
The exponent $\beta=0.662$ in Eq. (27) is also in good agreement with
the theoretical predictions $\beta_{\ast}=0.70\pm 0.02$, see
Ref. \cite{Str78}, and $\beta_{\ast}=0.72$, see Ref. \cite{Car89}.
Our estimate for the critical loading is confirmed by observations
depicted in Figures~4A and 4B, which reveal that transition from
the DC regime to the NR regime occurs at the degree of loading, $\phi$,
belonging to the interval between 30 and 50 phr.

The exponent $\kappa$ is plotted versus the degree of loading $\phi$
in Figure~5D, which demonstrates that the parameter $\kappa$
(which characterizes the rate of breakage for long chains) remains
constant for the NR regime.

To assess the effect of temperature $T$ on the distribution of long
chains, we fit experimental data for the CB filled bromobutyl rubber
in dynamic tensile tests at the temperatures $T=25$, $T=50$
and $T=100$~$^{\circ}$C.
The amplitude of strain oscillations $\Delta=0.01$ corresponds to the
regime of rupture for the filler network.
Figures~6A and 6B demonstrate fair agreement between observations and
results of numerical analysis with $N=40$, $N_{1}=80$ and $\kappa=16.0$.
In matching experimental data at various temperatures,
the parameter $\Gamma_{0}$ is treated as an adjustable parameter.
Surprisingly, it is found that its value
that ensures the best fit of experimental data
is independent of temperature and equals $\Gamma_{0}=0.0088$ s$^{-1}$.

The standard deviation of the number of strands, $\Sigma$, and the elastic
modulus $K$ are depicted in Figures~6C and 6D versus the
degree of undercooling, $\Delta T$.
Observations are approximated by the functions (22) used to match
experimental data for the CB filled natural rubber in the DC regime.
Comparison of Figure~2B with Figures~6C and 6D implies that the materials
under consideration demonstrate different behavior for the DC and NR
regimes.
In the regime of disjoint clusters, the growth of temperature
results in an increase in $\Sigma$ and a decrease in $K$.
On the contrary, in the regime of network rupture, the standard
deviation of the number of strands decreases and the elastic modulus
increases with temperature.
The rates of changes in these parameters increase with the volume
fraction of filler which may serve as a measure of connectivity
for the network.
The difference in the material response observed in these figures
is in agreement with the proposed scenario for mechanically induced
changes in the micro-structure of filled elastomers.

Finally, we approximate observations for styrene-butadiene
rubber filled by carbon black and polymeric particles in dynamic
shear tests with the strain amplitude $\Delta=0.002$
at the temperature $T=-30$~$^{\circ}$C.
A detailed description of specimens and the experimental procedure
is found in Ref. \cite{VUH99}.
Our aim is to assess the influence of the filler material on adjustable
parameters in the constitutive equations.

Figures~7A and 7B demonstrate fair agreement between experimental data
and results of numerical simulation with $N=40$, $N_{1}=80$,
$\Gamma_{0}=0.0012$ s$^{-1}$ and $\kappa=4.33$.
The quantities $\Sigma$ and $K$ are depicted versus the degree of loading
$\phi$ in Figures~7C and 7D.
Observations are fairly well approximated by the functions
\begin{equation}
\log \Sigma=C_{0}+C_{1}\phi,
\qquad
\log K=D_{0}-D_{1}\phi,
\end{equation}
where the constants $C_{i}$ and $D_{i}$ are determined by the least-squares
technique.
Figure~7 reveals transition from the DC regime of micro-structural changes
at small concentrations of filler to the NR regime at large degrees of
loading.
The critical concentration reads $\phi_{\rm cr}\approx 40$ phr
for CB particles and $\phi_{\rm cr}\approx 30$ phr for polymeric particles.
For the DC regime, the standard deviation of the number of strands, $\Sigma$,
slowly changes with $\phi$ (increases for the polymeric filler
and decreases for the CB filler, in agreement with data depicted
in Figure~4A), whereas the elastic modulus $K$ increases with $\phi$
for both kinds of filler (in agreement with observations presented in
Figure~4A).
For the NR regime, the quantity $\Sigma$ increases
and the modulus $K$ decreases with $\phi$ for both kinds of filler.
It is worth noting that the rates of changes in $\Sigma$ and $K$
with the degree of loading are similar for CB and polymeric particles.
This means that in the regime of disintegration of the filler network,
the viscoelastic response of filled elastomers is mainly determined
by the topological properties of the network and it rather weakly
depends on physical properties of the reinforcement.

\section{Concluding remarks}

Constitutive equations have been derived for the mechanical response
of particle-reinforced elastomers at small strains.
The model is applied to describe the Payne effect: changes in the dynamic
moduli of filled rubbery polymers with an increase in the amplitude of
strain-oscillations.
The viscoelastic behavior of reinforced rubbers demonstrates
two different regimes associated with (i) transformations of bound rubber
in the close vicinity of filler particles and their clusters
and (ii) fracture of the filler network and its disintegration into
disjoint clusters of particles.
The critical concentration of filler corresponding to the transition
from the DC regime to the NR regime is in good agreement with
theoretical predictions by the percolation theory, as well as with
experimental data for dc conductivity.

The following conclusions are drawn:
\begin{enumerate}
\item
Constitutive equations correctly predict observations in
dynamic mechanical tests on filled elastomers
at various temperatures, filler concentrations,
amplitudes and frequencies of oscillations.

\item
Changes in conditions of mechanical experiments
result in changes in adjustable parameters of the model
which are adequately described within the concepts of bound
and occluded rubber and which are in agreement with the proposed
scenario for transition from the DC regime to the NR regime
of deformation.
\end{enumerate}
The work is confined to the effects of temperature, filler concentration
and amplitude and frequency of strain oscillations
on the viscoelastic response of filled rubbers.
Some important questions, however, remain beyond the scope of
this study.
They include, in particular, (i) the influence of the shape of
particles, the distribution of their sizes \cite{CS99},
and the presence of coupling agents \cite{MTD93}
on the rheology of reinforced elastomers,
(ii) changes in the glass transition temperature of filled rubbery
polymers caused by polymer--filler interactions,
as well as (iii) the effect of mechanical loading
on electrical conductivity of reinforced elastomers \cite{FCB99,FHB01}.
These issues will be the subject of a subsequent publication.

\acknowledgments
Partial support from the European Commission under contract
No. G1RD--CT--1999--00085 is gratefully acknowledged.
AD acknowledges financial support by the Israeli Ministry
of Science through grant 1202--1--00.

\newpage \baselineskip = .5\baselineskip  

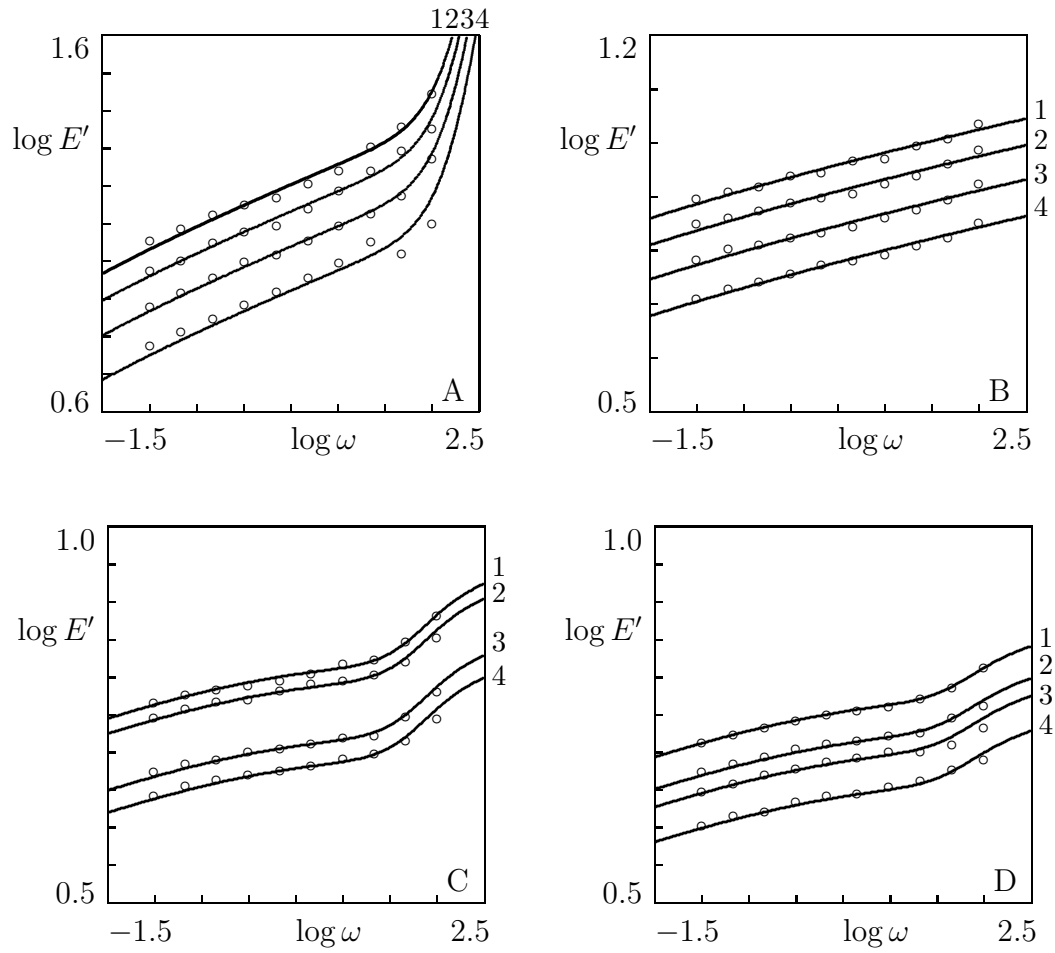
\begin{figure}[tbp]
\caption{The storage modulus $E^{\prime}$ MPa versus
the frequency of oscillations $\omega$ Hz for CB filled
natural rubber in a tensile dynamic test with the strain amplitude
$\Delta$ at a temperature $T$~K.
Circles: experimental data \protect{\cite{Lio98}}.
Solid lines: results of numerical simulation.
A: $T=253$; B: $T=296$; C: $T=333$; D: $T=373$.
Curve~1: $\Delta=0.006$;
curve~2: $\Delta=0.011$;
curve~3: $\Delta=0.028$;
curve~4: $\Delta=0.056$}
\end{figure}

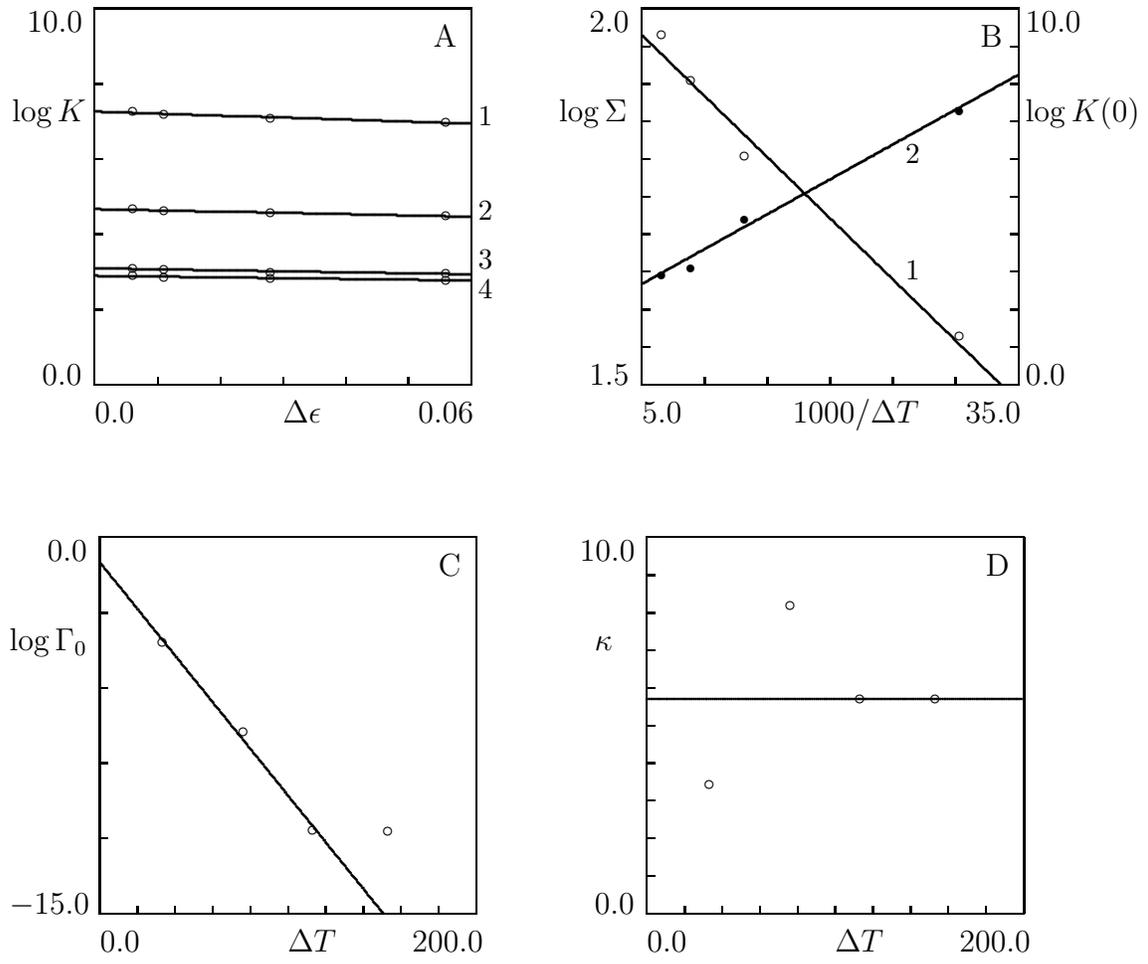
\begin{figure}[tbp]
\caption{A: The elastic modulus $K$ versus the amplitude of
oscillations $\Delta$ at a temperature $T$~K.
Solid lines: approximation of the experimental data by Eq. (21).
Curve~1: $T=253$, $a_{0}=7.2672$, $a_{1}=5.3129$;
curve~2: $T=296$, $a_{0}=4.6737$, $a_{1}=3.4019$;
curve~3: $T=333$, $a_{0}=3.0929$, $a_{1}=2.5039$;
curve~4: $T=373$, $a_{0}=2.8935$, $a_{1}=1.9944$.
B: The standard deviation of the number of strands $\Sigma$
(unfilled circles) and the initial modulus $K(0)$
(filled circles) versus the degree of undercooling $\Delta T$~K.
Solid lines: approximation of the experimental data by Eq. (22).
Curve~1: $b_{0}=2.0465$, $b_{1}=16.246$;
curve~2: $c_{0}=1.7553$, $c_{1}=185.33$.
C: The rate of breakage $\Gamma_{0}$ s$^{-1}$
versus the degree of undercooling $\Delta T$~K.
Solid line: approximation of the experimental data by Eq. (23)
with $d_{0}=-0.9827$ and $d_{1}=9.3009\times 10^{-2}$.
D: The coefficient $\kappa$ versus the degree of undercooling
$\Delta T$~K.
Solid line: the average value $\kappa=5.7$.
Symbols: treatment of observations for CB filled natural rubber
\protect{\cite{Lio98}}}
\end{figure}

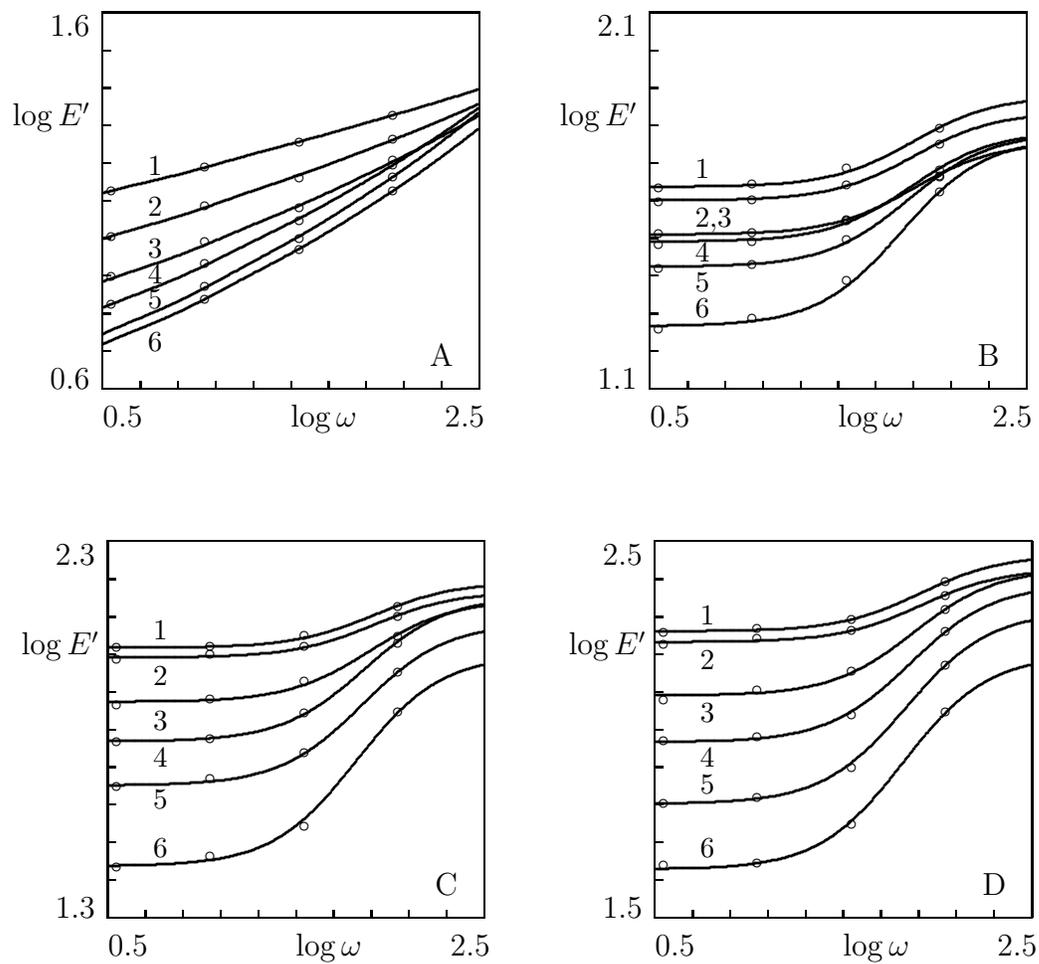
\begin{figure}[tbp]
\caption{The storage modulus $E^{\prime}$ MPa versus
the frequency of oscillations $\omega$ Hz for CB filled
bromobutyl rubber in a tensile dynamic test with the strain
amplitude $\Delta$ at room temperature.
A: $\phi=30$; B: $\phi=50$; C: $\phi=70$; D: $\phi=100$.
Circles: experimental data \protect{\cite{DKT91}}.
Solid lines: results of numerical simulation.
Curve~1: $\Delta=0.0003$;
curve~2: $\Delta=0.0011$;
curve~3: $\Delta=0.0033$;
curve~4: $\Delta=0.0048$;
curve~5: $\Delta=0.0077$;
curve~6: $\Delta=0.0122$}
\end{figure}

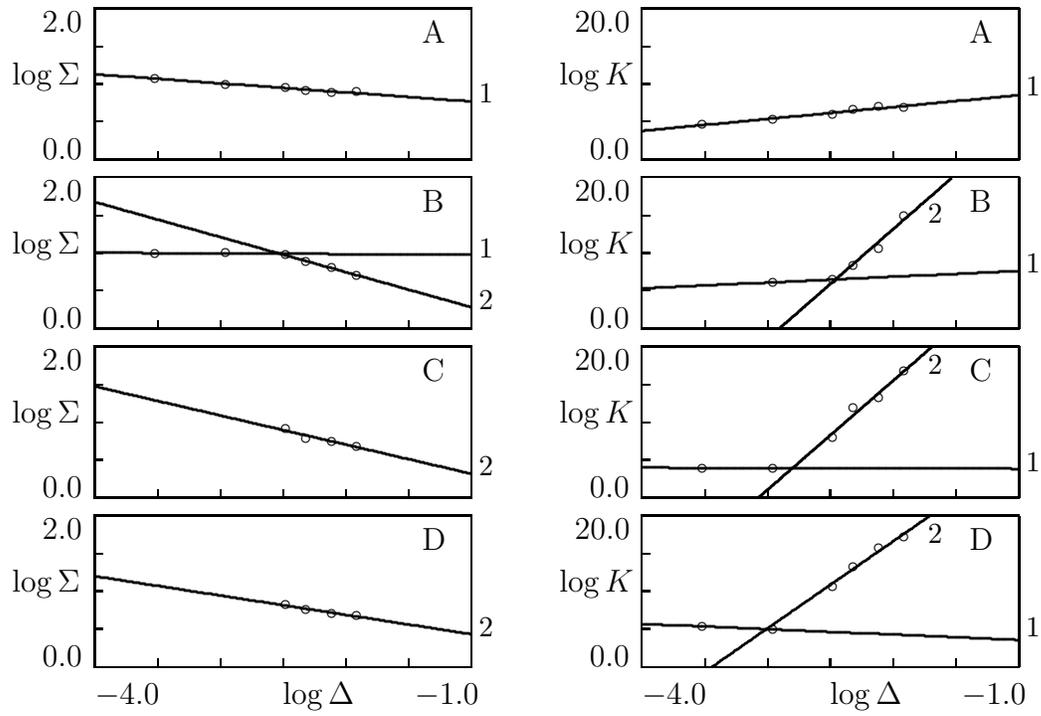
\begin{figure}[tbp]
\caption{The standard deviation of the number of strands
$\Sigma$ and the elastic modulus $K$ MPa
versus the amplitude of oscillations $\Delta$
for CB filled bromobutyl rubber.
Circles: treatment of observations \protect{\cite{DKT91}}.
Solid lines: approximation of experimental data by Eq. (24).
A: $\phi=30$,
Curves~1: $A_{0}=0.6474$, $A_{1}=0.1208$;
$B_{0}=10.0943$, $B_{1}=1.5797$;
B: $\phi=50$,
Curves~1: $A_{0}=0.9702$, $A_{1}=0.0084$,
$B_{0}=8.4060$, $B_{1}=0.7724$,
curves~2: $A_{0}=-0.1828$, $A_{1}=0.4658$,
$B_{0}=42.2894$, $B_{1}=14.5677$;
C: $\phi=70$,
Curve~1: $B_{0}=3.7590$, $B_{1}=-0.0458$,
curves~2: $A_{0}=-0.0706$, $A_{1}=0.3871$,
$B_{0}=44.4787$, $B_{1}=14.4917$;
D: $\phi=100$;
Curve~1: $B_{0}=2.8452$, $B_{1}=-0.7079$;
curves~2: $A_{0}=0.1749$, $A_{1}=0.2567$;
$B_{0}=39.7411$, $B_{1}=11.5563$}
\end{figure}

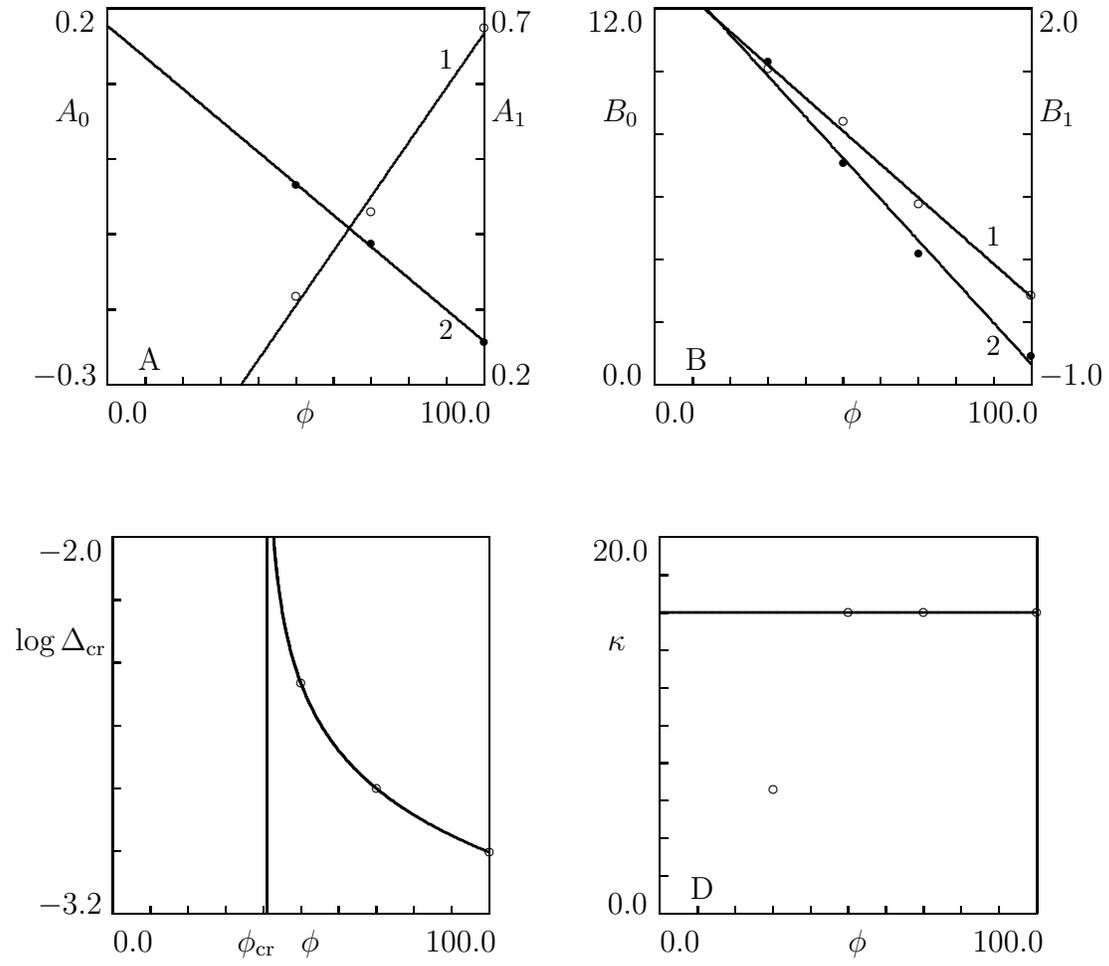
\begin{figure}[tbp]
\caption{A: The parameters $A_{0}$ (unfilled circles) and $A_{1}$
(filled circles) versus the degree of loading $\phi$ phr.
Solid lines: approximation of the experimental data by Eq. (25).
Curve~1: $A_{00}=-0.5568$, $A_{01}=7.2353\times 10^{-3}$;
curve~2: $A_{10}=0.6775$, $A_{11}=4.1950\times 10^{-3}$.
B: The parameters $B_{0}$ (unfilled circles) and $B_{1}$
(filled circles) versus $\phi$ phr.
Solid lines: approximation of the experimental data by Eq. (26).
Curve~1: $B_{00}=13.3942$, $B_{01}=0.1059$;
curve~2: $B_{10}=2.4530$, $B_{11}=0.0329$.
C: The critical amplitude of oscillations $\Delta\epsilon_{\rm cr}$
versus $\phi$ phr.
Solid line: approximation of the experimental data by Eq. (27)
with $C=1.4711\times 10^{-2}$, $\beta=0.6616$
and $\phi_{\rm cr}=40.99$ phr.
D: The parameter $\kappa$ versus $\phi$ phr.
Solid line: the limiting value $\kappa=16.0$.
Symbols: treatment of observations for CB filled bromobutyl
rubber at room temperature \protect{\cite{DKT91}}}
\end{figure}

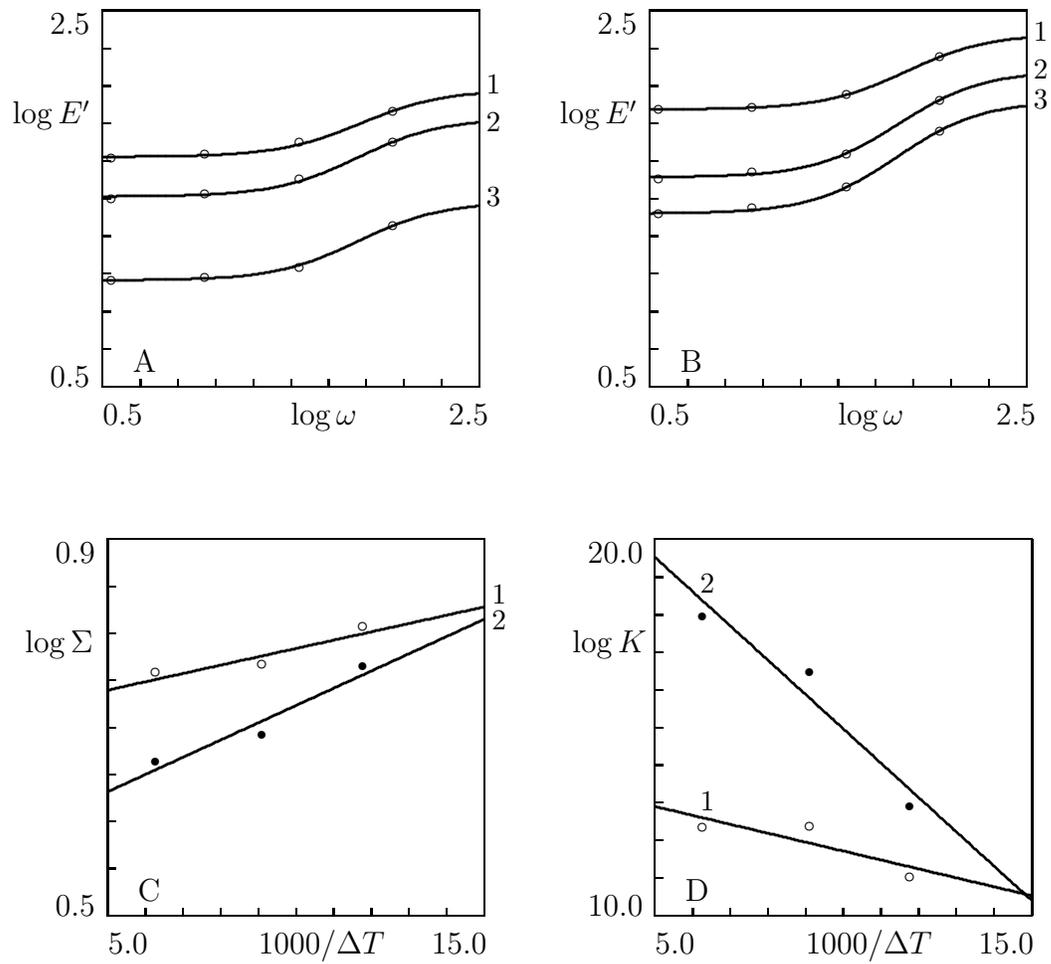
\begin{figure}[tbp]
\caption{A and B: The storage modulus $E^{\prime}$ MPa versus
the frequency of oscillations $\omega$ Hz for CB filled
bromobutyl rubber in a tensile dynamic test with
the strain amplitude $\Delta=0.01$ at a temperature $T$~$^{\circ}$C.
Symbols: experimental data \protect{\cite{DKT91}}.
Solid lines: results of numerical simulation.
A: $\phi=60$ phr, B: $\phi=100$ phr.
Curves~1: $T=25$;
curves~2: $T=50$;
curves~3: $T=100$.
C: The standard deviation of the number of strands $\Sigma$
versus the degree of undercooling $\Delta T$~K.
Symbols: treatment of observations.
Solid lines: approximation of the experimental data by Eq. (22).
Curve~1: $\phi=60$ phr, $b_{0}=0.6956$, $b_{1}=-8.8493$;
curve~2: $\phi=100$ phr, $b_{0}=0.5400$, $b_{1}=-18.337$.
D: The elastic modulus $K$ MPa versus the degree
of undercooling $\Delta T$~K.
Symbols: treatment of observations.
Solid lines: approximation of the experimental data by Eq. (22).
Curve~1: $\phi=60$ phr, $c_{0}=14.084$, $c_{1}=-236.96$;
curve~2: $\phi=100$ phr, $c_{0}=24.086$, $c_{1}=-911.59$}
\end{figure}

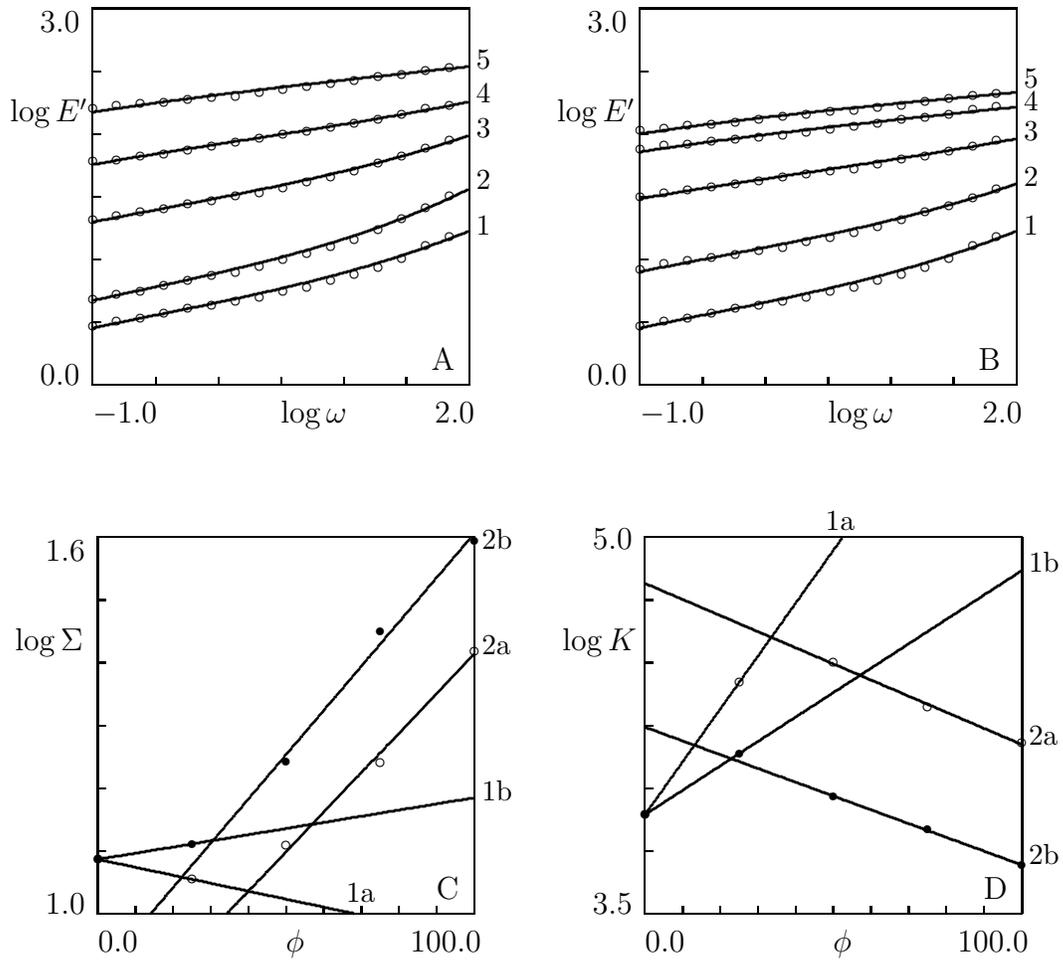
\begin{figure}[tbp]
\caption{A and B: The storage modulus $E^{\prime}$ MPa versus
the frequency of oscillations $\omega$ rad/s for filled
styrene-butadiene rubber (SBR--1500) with a degree of loading $\phi$ phr
in a tensile dynamic test with the strain amplitude
$\Delta=0.002$ at the temperature $T=-30$~$^{\circ}$C.
A: CB filled elastomer;
B: polymeric filled elastomer.
Circles: experimental data \protect{\cite{VUH99}}.
Solid lines: results of numerical simulation.
Curve~1: $\phi=0$;
curve~2: $\phi=25$;
curve~3: $\phi=50$;
curve~4: $\phi=75$;
curve~5: $\phi=100$.
C: The standard deviation of the number of strands $\Sigma$
versus the degree of loading $\phi$ phr.
Solid lines: approximation of the experimental data by Eq. (28).
Curve~1a: $C_{0}=1.0864$, $C_{1}=-0.0013$;
curve~2a: $C_{0}=0.7832$, $C_{1}=0.0063$;
curve~1b: $C_{0}=1.0864$, $C_{1}=-0.0010$;
curve~2b: $C_{0}=0.9003$, $C_{1}=0.0071$.
D: The elastic modulus $K$ MPa versus $\phi$ phr.
Solid lines: approximation of the experimental data by Eq. (28).
Curve~1a: $D_{0}=3.8947$, $D_{1}=-0.0211$;
curve~2a: $D_{0}=4.8175$, $D_{1}=0.0064$;
curve~1b: $D_{0}=3.8947$, $D_{1}=-0.0097$;
curve~2b: $D_{0}=4.2435$, $D_{1}=0.0055$.
Symbols: treatment of observations \protect{\cite{VUH99}}.
Unfilled circles: CB filler; filled circles: polymeric filler}
\end{figure}
\setcounter{figure}{0}
\newpage

\setlength{\unitlength}{0.5 mm}
\begin{figure}[t]
\begin{center}

\end{center}
\vspace*{10 mm}

\caption{}
\end{figure}
\newpage

\begin{figure}[t]
\begin{center}
%
\end{center}
\vspace*{10 mm}

\caption{}
\end{figure}
\newpage

\begin{figure}[t]
\begin{center}
%
\end{center}
\vspace*{10 mm}

\caption{}
\end{figure}
\newpage

\begin{figure}[t]
\begin{center}
%
\end{center}
\vspace*{10 mm}

\caption{}
\end{figure}
\end{document}